\documentclass{modified}

\begin{document}

\title{Code for RXMS Data Analysis}

\author{\footnotesize E.V.R.CHAN}

\address{ University of Washington,Box 351560\\
Seattle, Washington, 98195-2420, United States.\\
evr@u.washington.edu}

\maketitle

\begin{abstract}
The source code for resonant x-ray magnetic scattering 
(RXMS) data analysis is available.
It includes measures of the radial variation 
of intensity  and ability to extract azimuthal angular
variation of intensity within the transmission geometry.
If you need assistance to modify my code please 
email.
\end{abstract}


Due to recen$t^{1-5}$ interest in the azimuthal angle 
dependence of the RXMS intensities, I am making my
programs available (free for non-commercial applications).  
The code calculates radial and angular variation of intensity 
for transmission geometry and will need modification   
for your experimental arrangement if you should choose 
to utilize it in some manner.  By 
uncompressing the uploaded files with 
the programs uudecode and then
tar, you should find four text files.  These are the code
for producing the four figures in my previou$s^6$ 
publication. In the comments are additional code for 
very large color figures displayed on my webpage 
http://staff.washington.edu/$\sim$evr/Gallery.html  
involving fourier transform, inverse fourier,  
autocorrelation,etc.  A code fragment follows: 

\small  
\begin{verbatim}
%auto-ignore 

%%% From evr@u.washington.edu Thu Jul 10 14:59:51 2003
%%% Date: Thu, 10 Jul 2003 14:59:42 -0700 (PDT)
%%% From: evr evr@u.washington.edu
%%% To: "Allen, Rob" RAllen@roperscientific.com
%%% Cc: E. Chan evr@u.washington.edu
%%% Subject: RE: batch converting spe to tiff

%%% Subject: SPEfiles2Matlab.m
%%%  ~~~~~~~~~~~~~~~~~~~~~~~~~~~~
%%% reads Princeton camera *.SPE files into  Matlab
%%%      comments to  evr@u.washington.edu
%%% use at your own risk, may contain errors
%% fid=fopen('nameOfFile.SPE','r');
%% header=fread(fid,2050,'uint16');%4100bytes/2
%% ImMat=fread(fid,1024*1024,'uint16');
%% Z=reshape(ImMat,1024,1024);
%% fclose(fid);
%% Z=double(Z);
%% [X,Y]=meshgrid(1:1024,1:1024);
%% mesh(X,Y,Z); %display 3D
%%% ~~~~~~~~~~~~~~~~~~~~~~~~~~~~~~~~~~~~

%%%  On Wed, 9 Jul 2003, Allen, Rob wrote:

%%% ) can you tell me how you were able to do this, 
%%% ) so that i will be able to help others with 
%%% ) the same question?  thanks.
%%% )
%%%  )  ) -----Original Message-----
%%%  )  ) From: evr [mailto:evr@u.washington.edu]
%%%  )  ) Sent: Tuesday, July 08, 2003 10:07 PM
%%%  )  ) To: Allen, Rob
%%%  )  ) Cc: E. Chan
%%%  )  ) Subject: Re: batch converting spe to tiff
%%%  )  )
%%%  )  ) Thanks for all your advice.  
%%%  )  ) We can read the spe file into Matlab.
%%%  )  )

\end{verbatim}


\begin{thebibliography}{0}
\bibitem{1}
J.Beradkar and N.M.Kabachnik,
{\it J. Phys. B} {\bf 38}, 23 (2005).

\bibitem{2}
D. Bruns, B. Buchner, U. Gebhardt, S. Kiele, P. Reutler 
and A. Revcolevschi,
 {\it Phys. Rev. B} {\bf 69}, 104413 (2004).                          

\bibitem{3}
T. Matsumura, D. Okuyama, N. Oumi, K. Hirota, H. Nakao, 
Y. Murakami and Y. Wakabayashi,  
eprint cond-mat0411532.

\bibitem{4}
S. Ji, C. Song, J. Koo, K. B. Lee, Y. J. Park, J. Y. Kim, 
J.-H. Park, H. J. Shin, J. S. Rhyee, B. H. Oh and B. K. Cho, 
{\it Phys. Rev. Lett.} {\bf 91}, 257205 (2003).

\bibitem{5}
T. Nagao and J. Igarashi,
{\it J. Phys. Soc. Japan} {\bf 72}, 
2381 (2003). 

\bibitem{6}
E.V.R.Chan,eprint cond-mat0404371.


\end{thebibliography}
\end{document}